\documentclass[twocolumn,tighten,aps,epsfig]{revtex4}

\begin{document}
 
\title{Deuteron $NN^*$(1440) components from a chiral quark model}
 
\author{B. Juli\'a-D\'\i az,$^{1}$ D.R. Entem,$^{1,2}$ 
A. Valcarce,$^{1}$ and F. Fern\'andez$^{1}$ }
\address{$^1$ Grupo de F\' \i sica Nuclear,
Universidad de Salamanca, E-37008 Salamanca, Spain}
\address{$^2$ Department of Physics, University of 
Idaho, Moscow, Idaho 83844}

\begin{abstract}
We present a nonrelativistic coupled-channel calculation 
of the deuteron structure
including $\Delta\Delta$ and $NN^*$(1440) channels, besides
the standard $NN$ $S$ and $D$-wave components.
All the necessary building blocks to perform the
calculation
have been obtained from the same 
underlying quark model.
The calculated $NN^*(1440)$ probabilities
find support in the explanation given to different deuteron
reactions. 
\end{abstract}

\maketitle
\noindent
Since the discovery of nucleon structure, nucleon resonances
have attracted considerable attention from theorists and
experimentalists. 
An important effort was made to choose 
and design experiments that could probe the presence of
resonance configurations in different nuclear systems. 
The basic idea
lies in the observation that a small fraction of the nucleons
will be internally excited and therefore present as virtual 
resonances in every nucleus. This may happen even at low energies
due to the possibility of exciting internal nucleon degrees of freedom
according to the process 
$ NN \to NN^* \,\, ({\rm or} \,\, N^*N^*) \to NN$ 
involving intermediate $N^*$'s. As a consequence, the many
nucleon wave function should be supplemented by configurations involving
one or several nucleons in an excited baryon resonance state. 

If the virtual $N^*$'s exist in bound nuclear states, one expects
them to play an important role already in the bound two-nucleon
system, the deuteron \cite{WEAR}.
The most prominent low-lying even-parity 
nucleon resonances are the $P_{33}$, the
$\Delta(1232)$, and the $P_{11}$, the $N^*(1440)$ or Roper resonance.
The $N^*$'s contribute predominantly to the nucleon short-range
correlations and enhance the high momentum components of the nuclear
two-particle density. Being the deuteron isoscalar, the energetically
lowest state $N\Delta$ is forbidden, and therefore the
$\Delta\Delta$ and $NN^*(1440)$ components would be the relevant
nonnucleonic configurations.
The admixture probabilities of these {\it exotic} states 
are small due to the low nuclear density and the rather high
resonance-nucleon mass difference. Nonetheless, they have been
advocated long ago to understand elastic proton-deuteron backward
scattering at energies above pion threshold \cite{KEKI}
or the angular
distribution of deuteron photodisintegration at energies 
above $E_\gamma =$ 100 MeV \cite{HADJ}. 
Recent calculations have renewed the interest on
these nonnucleonic components  as they could be
indirectly observed in several reactions as for example
antiproton-deuteron annihilation \cite{DEKL}, subthreshold
antiproton production \cite{DOLR} or $pd \to dp$
processes \cite{UZIK}.
Although the evidence for resonance
configurations in the deuteron from such processes is indirect, is
suggestive and encouraging.

The treatment of nucleon resonances in the nuclear wave function 
can be done in different ways.
One possibility has consisted on keeping 
only nucleons (and no resonances) in the nuclear
wave function and using effective operators. 
However, it would be surprising if such an
{\it ad hoc} procedure 
would phenomenologically account for any detailed 
internal dynamics, because in the hard-core region two nucleons are 
likely to excite each other and thus mutually probe their internal
degrees of freedom.
If one wants to turn the attention to the virtual
contribution from these nucleon excited states, one
has to include explicitly $N^*$ transition potentials
in a coupled channel calculation. 

When performing a coupled channel calculation for the deuteron
based on effective baryon-baryon potentials, a problem immediately
arises. If one uses for the 
nucleon-nucleon ($NN$) channel an effective  potential which is
fitted to the $NN$ scattering data, 
it will already include 
contributions from intermediate $N^*$'s, and thus
one would obtain too much attraction at medium range. Therefore one
has to modify the normal nucleon-nucleon potential and weaken the
intermediate range attraction in order to account for the additional
attraction from the explicit dispersion contribution to the
potential with intermediate $N^*$'s. Such a procedure usually introduces
an unwanted model dependence on the results obtained.

There are multiple examples in the literature of 
these type of calculations.
Haapakoski and Saarela \cite{HASA} studied $\Delta\Delta$ 
components on the deuteron 
changing in an {\it ad hoc} manner
the intermediate range attraction of the central Reid soft-core
potential until the deuteron 
binding energy was fit to the experimental
value. The tensor force was not changed. Arenh\"ovel,
Danos and Williams \cite{ARDW} studied $NN^*(1440)$
configurations using perturbation 
theory with the one-pion-exchange (OPE) potential and 
Hulth\'en $NN$ deuteron wave functions.
The singular nature of the potentials required a cutoff factor
and the results were found to be rather sensitive to the cutoff chosen.
Weber \cite{WEBE} and Nath and Weber \cite{NAWE} 
calculated $NN^*$ components by means
of an OPE potential in momentum space. 
The resulting probabilities were again found to be
rather dependent on the high momentum suppression factor.
Finally, Rost \cite{ROST} performed
a calculation where the existence of resonance
components is taken into account by a modification of the
intermediate attraction of the Reid hard-core potential.
As a consequence, the study of nonnucleonic 
configurations in the deuteron based on standard
meson-exchange $NN$ potentials depends on two
basic assumptions: on one hand, the hypothesis
done to modify the intermediate
range attraction of the $NN$ interaction, on the
other hand, the specific transition potential to the
resonance configurations used. 

During the last decade 
an important effort has been devoted to the
qualitative and quantitative understanding of the 
nucleon-nucleon interaction based on quark
degrees of freedom. The first attempts of the early 80's based on the 
one-gluon-exchange (OGE) force \cite{OKYA,FAFL}
evolved to the so-called 
hybrid quark models \cite{KUNO,FEVS,GOLD}. 
The original OGE was supplemented  by
Goldstone-boson exchanges at the level of quarks and, thus,
fully quark-model based $NN$ potentials were obtained.
The OGE provided the first explanation of the repulsive
core of the $NN$ interaction. Although it is known that 
quark-antisymmetry on the Goldstone-boson exchanges induces the same 
effect \cite{VAFG}, the OGE 
still results crucial to regularize the short-range 
part of the Goldstone-boson exchange tensor force.
The pseudoscalar-boson exchange 
tensor coupling is fundamental to reproduce the $NN$ phase shifts,
and, finally, the scalar-boson exchange provides
the necessary medium-range attraction. The two-baryon problem 
based on quark-quark interactions is solved by means
of the resonating group method (RGM) in a coupled-channel
scheme considering usually
$NN$, $\Delta \Delta$ and hidden 
color-hidden color components. The importance of
particular configurations,
like $N\Delta$, to describe  
the low orbital angular momentum partial waves 
has been recently demonstrated \cite{VAFF}. 
Such calculations present
the great advantage that any baryon-baryon
interaction can be determined, once the $NN$ potential has been fixed,
in a completely parameter-free way. Therefore,
the quark-model framework
provides an adequate scheme to study nonnucleonic configurations
on the deuteron without the aforementioned uncertainties
appearing in meson-exchange models.

In the present work we want to focus 
on the influence of the most important 
nonnucleonic channels on the deuteron properties. In order 
to have a consistent 
calculation we will make use of baryonic potentials 
constructed from the same underlying quark-quark interaction
potential so that no parameters are fitted independently 
for the different channels. The quark-model parameters 
have been previously determined to describe the deuteron binding 
energy and $S$-wave $NN$ phase shifts 
with enough accuracy so that conclusions on the 
role played by the different channels, and thus, by the 
$N^*(1440)$ resonance or $\Delta$ can be inferred. 

Within the quark model framework,
this problem has already been partially undertaken
using different approximations. 
The possible effects of a bigger 
Hilbert space were considered in Ref. \cite{ZHBF} through 
the formation of six-quark bags at short-distances. 
Ref. \cite{KUNO} proposed an alternative
formulation in terms of quark-shell configurations, 
later on projected onto physical 
channels. Ref. \cite{GLKU} proposed an indirect
calculation where the nonnucleonic configurations were not
explicitly considered for the calculation of the two-body
system. As a consequence, these methods rely again on several
hypothesis that could hide physical conclusions.
In Ref. \cite{MAAM} the influence of $N$ and $\Delta$ 
resonances on the $NN$ interaction has been studied. The 
most significant contribution was obtained
from channels involving $N$ and $\Delta$ ground states,
although a quantitative calculation of the
nonnucleonic configurations was not performed.

Among the various constituent quark models 
already available in the literature, to have a reliable
prediction for the nonnucleonic components of the deuteron
one should require an accurate description of the
bound state and, at the same time, 
the $^3S_1$ and $^3D_1$ $NN$ scattering 
partial waves. For consistency
one should also require coherent results for the other
$NN$ $S$ wave, the $^1S_0$.
The model of Refs. \cite{FEVS,ENFV} fulfills the above 
requirements. 
The present study has been done by means of a 
Lippmann-Schwinger formulation of the
RGM equations in momentum space \cite{ENFV}.

In our calculation Eq. (20) of Ref. \cite{ENFV}  has been generalized 
to a coupled channel scheme.
It implies a modification of Eq. (24) of Ref. \cite{ENFV} with additional
terms which contain the different $NN \to NN^*$ couplings. 
The calculation of the transition potential 
has been simplified using
the Born-Oppenheimer approximation. We have 
carefully checked the quality of our approximation
in order to guarantee that our results will not
be influenced.  
It can be seen that the on-shell properties for the low 
angular momentum partial waves 
obtained by means of the Born-Oppenheimer 
method are of the same quality (within 5$\%$ for 
energies below 250 MeV)
as those obtained, using the same quark model Hamiltonian, 
but making use of the RGM \cite{BORG}.

The potential yields a fairly good reproduction
of the experimental data up to laboratory energies of 250 MeV.
In Table \ref{table0} we quote the quark model parameters, that
have been taken from Ref. \cite{ENFV}.
\begin{table}[h]
\caption{Quark-model parameters.}
\begin{tabular}{cccc}
& $m_q ({\rm MeV})$ & 313 &  \\
& $b ({\rm fm})$ & 0.518 &  \\
\colrule & $\alpha_s$ & 0.498  &  \\
& $\alpha_{ch}$ & 0.0258 &  \\
& $m_\sigma ({\rm fm^{-1}})$ & 3.538 &  \\
& $m_\pi ({\rm fm^{-1}})$ & 0.70 &  \\
& $\Lambda ({\rm fm^{-1}})$ & 4.30&
\end{tabular}
\label{table0}
\end{table}  
As can be seen, a fine tuning has been done
in order to consider the other nonnucleonic 
configurations.
Let us finally mention that the quark-quark interaction used
reproduces the even-parity baryon 
spectrum ($N$, $N^*(1440)$, $\Delta(1232)$,
$\Delta(1600)$,..) in an exact Faddeev calculation \cite{BRIE}.

Within the above described framework we have calculated 
the deuteron binding energy and wave function making emphasis in
a simultaneous description of the $NN$ scattering phase shifts.
For the bound state problem, Eq. (20) of Ref. \cite{ENFV}
can be discretized in momentum space and written as
\begin{equation}
\sum_j ( H_{ij}- E  \delta_{ij}) \Psi_j = 0 \, ,
\label{eq1}
\end{equation}
where we have used a simplified notation and the indices
$i$ and $j$ run not only for all the discretization points 
but also for the different channels included in the
calculation. 
The nontrivial solutions are given by
the zeroes of the Fredholm determinant
\begin{equation}
| H_{ij}-E  \delta_{ij} | = 0 \, ,
\label{fredholm}
\end{equation}
being the values of $E$ that satisfy the previous equation
the energy of the bound states. Once the energies have been 
found the wave function can be easily calculated 
solving the linear problem of Eq. (\ref{eq1}).

We show in Table \ref{table1} the different configurations
and partial waves included 
in our calculation. 
\begin{table}[h]
\caption{Different channels and partial waves
considereded in our calculation. 
We give on the second column the energy difference (MeV)
with respect to the $NN$ system}
\begin{tabular}{ccccc}
 & $NN$&$^3S_1$ - $^3D_1$ & 0.0 & \\
 & $NN^*(1440)$&$^3S_1$ - $^3D_1$ & 501.0 & \\
 & $\Delta\Delta$&$^3S_1$ - $^3D_1$ - $^7D_1$ - $^7G_1$&586.0 & \\
\end{tabular}
\label{table1}
\end{table}                 
In Table \ref{table2}  we present the results obtained for  
the nonnucleonic probabilities and the static properties of
the deuteron. In all cases the deuteron binding energy is correctly
reproduced, being $E_d=-$2.2246 MeV. 
We have shown the results of a calculation
including only $NN$ components, including $NN$ and $\Delta\Delta$
configurations and finally the full calculation including also
$NN^*(1440)$ configurations. 
Among the allowed configurations, the $NN^*(1440)$
has not been
usually included in the deuteron calculations due to the
great uncertainty associated to the coupling constant
and cutoff parameters \cite{ARDW,ROST}. 
The first result we would like to emphasize
is the fact that the probability of $NN^*(1440)$ channels are 
smaller than the $\Delta\Delta$ ones. 
They do not show much influence on the static properties
of the deuteron as it seems to be case for the deuteron
form factors \cite{GROS}. However, these small components
find support in the explanation given in the literature
to some deuteron reactions \cite{DEKL,DOLR,UZIK}. 
Subthreshold antiproton
production in $d - p$ and $d - d$ reactions \cite{DOLR}, 
$pd \to dp$ reactions \cite{UZIK}, or
antiproton-deuteron annihilation at rest \cite{DEKL} are
compatible with small percentage of $NN^*(1440)$ in the 
deuteron wave function.

The prediction we obtain for the $NN^*(1440)$
probabilities, a larger component of the
$^3D_1(NN^*(1440))$ partial wave than the 
$^3S_1(NN^*(1440))$ partial wave,
agrees with the ordering obtained
by other calculations present in the literature \cite{ROST,MAAM}.
This can be understood if one takes into account that
the tensor coupling, which is the main responsible for the
presence of nonnucleonic components on the
deuteron wave function, is much stronger
for the $^3S_1 (NN) \to \! ^3D_1 (NN^*(1440))$ transition than 
for the $^3D_1 (NN) \to \! ^3S_1 (NN^*(1440))$ one, 
enhancing in this way the $D$-wave influence with 
respect to the $S$-wave component.
Regarding the absolute value of the probabilities, 
our results are a factor ten 
smaller than those reported on Ref. \cite{ROST}, where 
an estimation of 0.17$\%$ for the $NN^*(1440)$ 
configuration was obtained (0.06$\%$ for the $^3S_1$
and 0.11 $\%$ for the $^3D_1$ partial wave). 
The dependence of this
result on the hypothesis made and the deviation
from the results we obtain could be 
understood in the following way. The deuteron is calculated using
the Reid hard-core potential. When including $NN^*(1440)$
components, the channel coupling induces an attractive 
interaction on the $NN$ system, that needs to be subtracted
out. Such a subtraction was done by 
reducing the intermediate range attraction
of the central part of the Reid hard-core potential without
modifying the tensor part, as was done in Ref. \cite{HASA}
to calculate the probability of $\Delta\Delta$ components. 
As a consequence, in these type of 
calculations the strength of the tensor coupling to the
$NN^*(1440)$ state can be enhanced by decreasing the intermediate
range attraction in the $NN$ channel. The balance between these
two sources of attraction cannot be disentangled in a clearcut way.
This is a similar problem to the one arising in the $^1S_0 (NN)$ partial
wave when the coupling to the $N\Delta$ system was included \cite{GREE}.
The same attractive effect could be obtained by a central potential or 
a tensor coupling to a state with higher mass, being necessary
other observables to discriminate between 
the two processes \cite{GREE}.
This seems to be the reason of the much bigger probability for 
the $NN^*(1440)$ components in Ref. \cite{ROST}, that
on the other hand
showed a great
dependence on the choice of the $NN$ phenomenological potential.
In Ref. \cite{MAAM}, although the contribution of resonance
configurations has been included to study the nucleon-nucleon
system, there are no numerical predictions to compare with. 

There are other estimations on the literature.
The results of Ref. \cite{ARDW} are only of qualitative
interest. The pathological behavior of the transition
potential to resonance states was regularized by a 
cutoff factor that made the potential too weak at
small distances.
In Ref. \cite{GLKU} they 
study the effective numbers for different
resonance configurations on the deuteron making use
of baryon wave functions obtained from the diagonalization
of a quark-quark interaction containing gluon and
pion exchange in a harmonic oscillator basis including
up to 2$\hbar \omega$ excitations, and deuteron wave 
functions obtained from the Paris potential or a different
quark model approach. They obtain an upper limit
of $1\%$ for $\Delta\Delta$ components and $0.1\%$ for
$NN^*(1440)$ in agreement with the order of magnitude
and ordering of our results. 

To summarize, we have studied the
resonance structure of the deuteron 
using quark-model based interactions.
The $NN^*(1440)$ baryon resonance probabilities
in the deuteron, being smaller than the $\Delta\Delta$ ones,
are in agreement with indirect estimations obtained
from the analysis of processes like subthreshold 
antiproton production on $d-p$ collisions or $pd \to dp$ reactions. 

B.J. thanks Ministerio de Ciencia y Tecnolog\'{\i}a (Spain)
for finantial support.
This work has been partially funded by 
Ministerio de Ciencia y Tecnolog\'{\i}a (Spain) under Contract
No. BFM2001-3563, by Junta de Castilla y Le\'on under Contract No. 
SA-109/01, and by the Ram\'on Areces Foundation.

\begin{table}[h]
\caption{Deuteron wave function ($\%$)}
\begin{tabular}{cc|cc|cccc|ccc}
\multicolumn{2}{c}{$NN$}&\multicolumn{2}{c}{$NN^*$(1440)}&
\multicolumn{4}{c}{$\Delta\Delta$}&\multicolumn{3}{c}{}\\
$^3S_1$ & $^3D_1$ & $^3S_1$  & $^3D_1$ &  
$^3S_1$ & $^3D_1$ & $^7D_1$ &
$^7G_1$&$r_m$(fm)&$A_S$(fm$^{-1/2}$)&$\eta$\\
\hline
95.3780&4.6220&-&-&-&-&-&-&1.976&0.8895&0.0251\\
95.1989&4.5606&-&-&0.1064&0.0035&0.1243&0.0063&1.985&0.8941&0.0250\\
95.1885&4.5377& 0.0022&0.0148&0.1224&0.0036&0.1245&0.0063&1.985&0.8941&0.0250\\
\end{tabular}
\label{table2}
\end{table}

\end{document}